# Generalised ontological models based on processes

Tung Ten Yong

tytung2020@gmail.com

**Abstract**

In this paper, we explore realist models of quantum theory that does not fit into the standard definitions of ontological models. The models here go beyond standard definition of ontological models in the sense that quantum states do not correspond to distributions over the ontic state space and a system prepared in a quantum state is not in an ontic state. Instead, a system in a quantum state is always in a process – moving around in the ontic state space. Also, quantum measurement outcomes are not direct measurement of the ontic state, but depend probabilistically on the entire path the system takes during the measurement process. Consequently, we explain how, in our model, quantum states can neither be classified as ontic nor epistemic in the sense of knowledge about an underlying reality. In our model, quantum probabilities describes our (objective) knowledge about measurement outcomes. We also look at two hybrid models where either the preparation or measurement do follow the definitions in standard ontological models. Lastly, we propose a form of generalised ontological model that reduces to the standard PBR model when the underlying process reduces to a point in ontic space.

**Introduction**

Ever since the advent of quantum theory, there has been different interpretations about the meaning of the theory. For example, do the quantum states refer to the physical states of the observed system or do they refer to our knowledge/information/belief? Can the quantum probabilities be obtained from the properties of the system (and, possibly, together with the settings of measurement apparatus)? Whereas we have obtained certain results, in the form of no-go theorems, on questions about (non-)locality [1] and (non-)contextuality [2], questions about the above-mentioned questions have only recently received their own no-go theorems. In a series of papers [3, 4] (see also [5]), the questions of whether quantum states should be seen as representing physical states of the system, or as observer's information about the system, are being investigated in the form of *ontological models*.

Ontological models of quantum theory is the formalism that aims to captures the general structure of realistic theories of quantum theory. In ontological models, real physical states of the system, the *ontic states*, are assumed, and quantum states correspond to *probability distributions* over some set of ontic states. To those who hold an epistemic view of quantum states, it was hoped that an important feature of quantum theory – the non-distinguishability of distinct non-orthogonal states in a single shot experiment could be explained (even if partially) as overlapping probability distributions in ontological models. However, in [4], it was shown that (under several assumptions such as the independence of preparations) the models cannot recover quantum predictions unless the distributions for distinct quantum states are always disjoint.

In this paper, we would like to argue that, as a framework that aims to capture the possible realistic underpinnings of quantum theory, the standard form of ontological models are too restrictive in their definition. In the first section, we will review the standard definition of ontological models, followed by a discussion of why we think it is too restrictive. We then provide a simple ontological model that does not fit into the standard ontological models. Several implications are discussed in the section following that, where we also propose two hybrid ontological models. Lastly, we give a generalisation of the ontological framework for quantum theory that is process based.

**I. Ontological models**

An ontological model assumes that when a physical system is prepared in a quantum state $|\psi\rangle$, it is actually in an ontic state, $\lambda$, the real physical state of the system. $\lambda \in \Lambda$, where $\Lambda$ is the total set of ontic states.

In the model, the quantum state $|\psi\rangle$ corresponds to a probability distribution over $\Lambda$, $p(\lambda|\psi)$. The probability distribution represents our ignorance about which ontic state the system actually resides in, given that all we know about the system is that it is prepared in the quantum state $|\psi\rangle$. Therefore,

$p(\lambda|\psi) \geq 0$ and $\int p(\lambda|\psi)\, d\lambda = 1$

The model assumes that when a measurement is carried out, the probability of outcomes depends on the ontic state of the system and the measurement apparatus. Therefore, for measurement $M$, the outcome $E$ occurs according to some probability $p(E|\lambda, M)$. Thus:

$p(E|\lambda, M) \geq 0$ and $\sum_{\{E\}} p(E|\lambda, M) = 1$

For the model to reproduce quantum probabilities, the two probabilities above must satisfy

$$p(E|\psi, M) = \int p(\lambda|\psi)\, p(E|\lambda, M) d\lambda = |\langle E|\psi\rangle|^2$$

for all $\psi$ and $E$.

Note that for the above to hold, the following are assumed

$$p(\lambda|\psi, M) = p(\lambda|\psi) \qquad (1)$$

and

$$p(E|\lambda, \psi, M) = p(E|\lambda, M) \qquad (2)$$

Equation (1) is usually assumed in order to avoid retrocausality and superdeterminism. We will make this assumption in this paper. Equation (2) means that $\lambda$ contains more information (in fact, complete information) about the system, as compared to the information from knowing only $\psi$.

The above applies to one or more quantum systems. For two or more quantum systems one can make further assumptions about the system, such as independence of preparation for unentangled systems $\psi_1$ and $\psi_2$: $p(\lambda_1, \lambda_2|\psi_1, \psi_2) = p(\lambda_1|\psi_1)p(\lambda_2|\psi_2)$. We will not go into more than one quantum system in this paper.

## II. Limits of ontological models

A common assumption in the above ontological models is that when the system is prepared in a quantum state $|\psi\rangle$, it is actually in an ontic state $\lambda$.

We would like to argue that this requirement is too restrictive for ontological models of quantum theory. An analogy to thermodynamics is apt here. In thermodynamics, a system in static equilibrium state is, in fact, always traversing around some microstates compatible with the macrostate. Therefore, in ontological models, it would also be reasonable to relax the assumption that the system in a quantum state $|\psi\rangle$ resides in one fixed ontic state. As in thermodynamics, even though a system is described by a quantum state $|\psi\rangle$ at the quantum level, it might undergo certain *process* at the level of ontic states.

Another restrictive assumption in the current ontological models is related to how the measurements are modelled. As shown in section II, current ontological models represent measurements in the form $p(E|\lambda, M)$. However, this representation is closely related to the assumption mentioned above. Representing measurement as $p(E|\lambda, M)$ is reasonable if the system is in an ontic state $\lambda$ during the measurement process. It would also be reasonable if the measurement outcome only depends on the ontic state of the system *at a single moment* within the interval of measurement process.

But if the system actually traverses a set of ontic states during the measurement process, the probability of measurement outcomes might depend on the entire set, not just on one particular ontic state.

## III. A simple process-based model of quantum state and measurements

Now we present an ontological model that does not fit into the standard framework of ontological models. For the sake of simplicity we give a model that reproduces quantum probabilities of a spin-1/2 system.

First, we will define the ontic states. The state space comprises of different sets, each corresponds to a spin measurement in the direction $\theta$. Each set contains two elements, each corresponds to one measurement outcome in that direction. Therefore, we label the states as $\lambda_{\theta_\alpha}$, where $\alpha = -1, +1$. The state space $\Lambda = \{\lambda_{\theta_\alpha} : \theta$ for all directions, $\alpha = -1, +1\}$.

In this model, when the system is in any quantum state $|\psi\rangle$, it is actually in a process: it is traversing the ontic state space. Our assumption is that the system traverses *every* ontic state *except* for the one that corresponds to the quantum state orthogonal to $|\psi\rangle$. So, if the system is prepared in $|z+\rangle$, it is actually jumping around all ontic states except for $\lambda_{z-}$. We do not make any assumption about the dynamics of this underlying process, allowing the possibility of an entirely random process.

$|z+\rangle$:

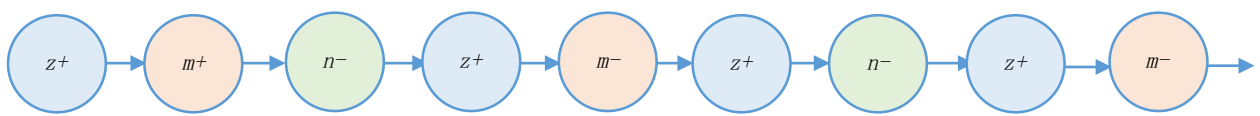

*Figure 1.* Example of a path traversed by the system prepared in $|z+\rangle$.

But, in order to reproduce the quantum probabilities, all we need is that the number of times a particular ontic state being traversed, $N_\psi(\theta_\alpha)$, follows the rule:

$$\frac{N_\psi(\theta_\alpha)}{\sum_\alpha N_\psi(\theta_\alpha)} = |\langle\theta_{\alpha_0}|\psi\rangle|^2$$

This is the only rule bounding the underlying process. Let's call this the *Relative Frequency Rule*.

For example, for a system in state $|z+\rangle$, the number of times $\lambda_{m_+}$ is being travelled, $N_{z+}(m_+)$, is given by:

$$\frac{N_{z+}(m_+)}{N_{z+}(m_+) + N_{z+}(m_-)} = |\langle m+|z+\rangle|^2$$

Since for a system in state $|z+\rangle$, it will not traverse $\lambda_{z-}$, we have $N_{z+}(z_-) = 0$ and recover the trivial probabilities:

$$\frac{N_{z+}(z_+)}{N_{z+}(z_+) + N_{z+}(z_-)} = \frac{N_{z+}(z_+)}{N_{z+}(z_+)} = 1 = |\langle z+|z+\rangle|^2$$

$$\frac{N_{z+}(z_-)}{N_{z+}(z_+) + N_{z+}(z_-)} = \frac{0}{N_{z+}(z_+)} = 0 = |\langle z-|z+\rangle|^2$$

Therefore, in this model we only set the number of times a certain ontic state is travelled, *relative* to the total number of times states in that same direction are travelled. It is the relative frequency within a single direction that is meaningful, while the absolute frequency over all directions is not restricted.

Now, we will describe quantum measurement in this model. In this model, measurement is fundamentally a process *in time*. What this means is that the (probability of) outcomes do not depend on the state of the world of just a single moment. In fact, they could be determined by the entire process occurring at the ontic level. This is the reason, as also pointed out in the previous section, why $p(E|\lambda, M)$ is not a well-defined expression in our model.

More specifically, in our model the measurement process takes a certain interval of time. Within this interval of time it examines the set of ontic state corresponding to the input quantum state $|\psi\rangle$. Let's say we are measuring spin in the $m$-direction and the incoming quantum state is $|z+\rangle$. An important characteristic of measurement process in this model is that it can only "see" the ontic states corresponding to this direction: $\lambda_{m_+}$ and $\lambda_{m_-}$. The apparatus will then *randomly* choose one of the ontic states visible to it. Since it is random, all $\lambda_{m_+}$ and $\lambda_{m_-}$ are equally likely to be chosen.

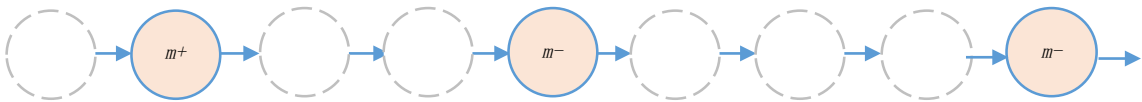

*Figure 2.* Ontic states that is visible by an apparatus measuring spin in the $m$-direction.

The probability of choosing $\lambda_{m_+}$, for example, is therefore given by

$$\frac{N_{z+}(m_+)}{N_{z+}(m_+) + N_{z+}(m_-)}$$

We also assume that *during this interval of measurement process, the relative frequency rule holds*.

By the relative frequency rule, this is equal to $|\langle m + | z + \rangle|^2$. Similarly, the probability of choosing $\lambda_{m_-}$ is

$$\frac{N_{z+}(m_-)}{N_{z+}(m_+) + N_{z+}(m_-)} = |\langle m - | z + \rangle|^2$$

In this way the measurement probabilities in the model recovers the quantum probabilities.

After choosing, for example, the state $\lambda_{m_+}$, the measurement will show the measurement result as + We also assume that, as a result of this interaction, the observed system will continue to traverse in the ontic state space, albeit in a different path now that is compatible with $|m +\rangle$. This assumption is required to recover the fact that an immediate measurement of spin in *m* direction will return outcome + with probability 1.

As one can see from this way of modelling measurement process, probabilities of outcomes do not just depend on any one ontic state, certainly not on one ontic state at a single moment of the measurement process. It depends on how frequent some states are travelled during the entire interval of measurement. But we could say more: the expression $p(\lambda|\psi)$ is also not well-defined. Let's take the example of $p(\lambda_{m_+}||z +\rangle)$. In our model, this could only mean

$$p(\lambda_{m_+}||z +\rangle) := \frac{N_{z+}(m_+)}{\sum_{\theta,\alpha} N_{z+}(\theta_\alpha)}$$

That is, count how many times the state $\lambda_{m+}$ is travelled, and divide it over the *total* number of ontic states travelled during the entire interval of measurement process. However, our model do not say anything about the total number of ontic states travelled during the interval, both $N_{z+}(m_+)$ and $N_{z+}(m_+) + N_{z+}(m_-)$ are arbitrary, as long as the relative frequency

$$\frac{N_{z+}(m_+)}{N_{z+}(m_+) + N_{z+}(m_-)}$$

is fixed.

Therefore, as we have seen from this simple model, both $p(\lambda|\psi)$ and $p(E|\lambda, M)$ are not well-defined, because the system undergoes an underlying process in any quantum state and measurement outcome do not depend on a single ontic state.

## IV. Implications

*Quantum states – neither ontic nor epistemic*

It is clear from the above discussions that in our model quantum state does not even satisfy the necessary condition for being epistemic, i.e. corresponding to a distribution over the ontic state space, as the expression $p(\lambda|\psi)$ is not well-defined.

However, quantum state is also not ontic in our model. Quantum state is ontic if system's ontic state determines its quantum state. However, this is not true in general in our model, for example, ontic state $\lambda_{m_+}$ lies in the paths corresponding to both $|z+\rangle$ and $|m+\rangle$. Knowing $\lambda_{m_+}$ does not tell us which quantum state the system is in (not even probabilistically).

The probability that appear in the model can be viewed as epistemic. Probability arise in the model at the stage of measurement process, where it *randomly* chooses ontic states that are visible to it. We can view this randomness as our complete lack of knowledge about the workings of how the ontic state is chosen, given that all we know is the input quantum state. Therefore, in this model, quantum state is epistemic about possible measurement outcomes, not about the underlying ontic states. Also, this knowledge is objective as anyone who knows about the process will assign the same probabilities to measurement outcomes, via the relative frequency rules.

*Dynamical model of the quantum probabilistic structure*

Unlike the standard ontological models, the model proposed in this paper is a dynamical model, in the sense that the (timeless) probabilistic structure of quantum theory has a dynamical origin in this model. Both the preparation and measurement of a quantum state involves processes occurring at the level of ontic states. Moreover, interesting features of quantum theory can be explained in this model by the properties of these processes:

a) **Orthogonality** results from certain corner of the ontic space not traversed by the system;

b) **Indistinguishability** of two quantum states $|\psi\rangle$ and $|\phi\rangle$ is due to the apparatus measuring $|\phi\rangle$ "sees" a system, which prepared in $|\psi\rangle$, traversing those ontic states that are correspond to $|\phi\rangle$.

*Hybrid models - Partially dynamical models*

Inspired by the model, we here proposed two models that are partially dynamical and partially probabilistic.

### a) Dynamical-Probabilistic Model (D-P model)

By this we mean models in which the preparation of a quantum state $\psi$ corresponds to a process over the ontic states; while measurement outcome is, as in standard ontological models, triggered probabilistically by ontic state of the system at a moment. It is easy to create a D-P model that reproduces orthogonality and indistinguishability.

Assume a minimal of three distinct ontic states, labelled as *a, b, c, d*. A system prepared in $|z+\rangle$ is traversing these states in the following pattern:

$$a, b, a, b, a, b, \ldots$$

A system prepared in $|m +\rangle$ traverses the following pattern:

$$b, c, b, c, b, c, \ldots$$

While a system prepared in $|z -\rangle$ traverses the pattern:

$$c, d, c, d, c, d, \ldots$$

Assume that measurement outcomes in this model are triggered by the ontic states at any single moment. We consider the simple situation where each ontic state triggers one and only one of the possible outcomes for any measurement setting.

To reproduce indistinguishability of these two quantum states we need a two-outcome measurement $M$. Define the detection rule as: If it finds the system to be in $a$ or $b$, $M$ gives outcome $m_1$; if it detects the system is in $c$ or $d$, $M$ gives outcome $m_2$. Therefore, both $|z +\rangle$ and $|m +\rangle$ could trigger outcome $m_1$ because they both contain ontic state $b$; similarly, both $|z -\rangle$ and $|m +\rangle$ could trigger outcome $m_2$ as they both contain ontic state $c$.

**Orthogonality:** If the input system is in either $|z +\rangle$ or $|z -\rangle$, an outcome of $m_1$ tells us that the system is in quantum state $|z +\rangle$ with certainty; while an outcome of $m_2$ tells us that the system is in quantum state $|z -\rangle$ with certainty. The two states are distinguishable in a one shot measurement $M$.

**Indistinguishability:** If the input system is in either $|z +\rangle$ or $|m +\rangle$, an outcome of $m_2$ tells us that the system is in quantum state $|m +\rangle$ with certainty. However, an outcome of $m_1$ does not tell us if the system was prepared in $|z +\rangle$ or in $|m +\rangle$, only that half of the time it is in $|z +\rangle$ and the half of the time it is in $|z +\rangle$. The two states are indistinguishable according to $M$. In fact, for two-outcome measurements, no detection rule could be assigned in such a way that $|z +\rangle$ or $|m +\rangle$ are distinguishable in a single shot measurement, as they both traverses the same ontic state $b$.

In fact, in this model one can define three different two-outcome measurement settings:

i) $a, b \rightarrow m_1 \,; c, d \rightarrow m_2$

ii) $b, c \rightarrow n_1 \,; d, a \rightarrow n_2$

iii) $b, d \rightarrow k_1 \,; a, c \rightarrow k_2$

This becomes analogous to Spekken's toy model [6] if the measurements above are identified as the z-measurement, y-measurement and x-measurement, respectively, of a spin-1/2 system, and the outcomes are either -1 or +1.

b) Probabilistic-Dynamic Model (P-D model)

One of the major reason that quantum state is taken to be epistemic is the indistinguishability of distinct non-orthogonal states in a single measurement. To explain this, distinct non-orthogonal quantum states correspond to overlapping distributions on the ontic space. Therefore when system's ontic state lies in the overlapping region, no single measurement is able to determine with certainty which quantum state it was prepared. However, overlapping seems like a good explanation only because our view of measurement is too restrictive. In standard ontological models measurements are represented by $p(E|\lambda, M)$, a probabilistic trigger of outcomes by the ontic state of the system. However, quantum measurements cannot be mere passive registrations of states without actively altering the states, or it will run afoul of the quantum uncertainties. Disturbance of the system is usually added into the ontological model in an ad hoc way, requiring it to happen after the outcome is registered.

In this P-D model, we simply assume that measurement first triggers the system to follow certain trajectory in the ontic space and then gives an outcome that depends probabilistically on the trajectory. In this way, outcome and disturbance have the same (dynamical) origin in the P-D model.

In this hybrid model, preparation is described, as usual, as a distribution $p(\lambda|\psi)$. However, measurement is described as distribution where outcome depends on paths in ontic space, $p(E|\mathcal{P}, M)$. The outcome probability depends on the entire path $\mathcal{P}$, not just on one ontic state.

This model is based on two rules:

a) $p(E_n|\mathcal{P}, M_n) > 0$ iff $\lambda_{E_n} \in \mathcal{P}$;

b) $\forall \mathcal{P} \in [\lambda_{E_n}; M_m]$,
   i) $\lambda_{E_n}, \lambda_{E_n'} \in \mathcal{P}$ $\quad (n \neq m)$
   ii) $\lambda_{E_n} \in \mathcal{P}, \lambda_{E_n'} \notin \mathcal{P}$ $\quad (n = m)$

where $M_n$ is measurement of spin in the $n$-direction and $E_n$ is an outcome of $M_n$. $\lambda_{E_n}$ is an ontic state that corresponds to outcome $E_n$. $E_n'$ is any outcome complement to $E_n$.

The first rule says that a measurement of $M_n$ gives outcome $E_n$ if certain ontic states lies on the path triggered by $M_n$. The second rule says that ontic states that trigger results complement to $E_n$ would lie on the path $[\lambda_{E_n}; M_m]$ unless the input quantum state is an eigenstate of $M_n$.

This model explains orthogonality and indistinguishability in the following way:

**Orthogonality:** Two quantum states are orthogonal if their distributions are non-overlapping and ontic states from their supports following non-intersecting trajectories in ontic space, triggered by some appropriate measurement.

Let the measurement gives outcome $m_1$ if the system passes through certain ontic states (those lying on one of the paths), and gives outcome $m_2$ if the system passes through some other states (that lie on the second path). This measurement therefore distinguishes these orthogonal quantum states in one shot.

**Indistinguishability:** Two quantum states are indistinguishable if, under the trigger of any measurement, ontic states in their support move along paths that intersect with each other at ontic states that gives nonzero probabilities to at least one outcome of that measurement.

Seeing in this light, our model in section IV is a D-D model (Dynamic-Dynamic model), where both preparation and measurement are both dependent on the entire process and neither are distributions over the ontic space.

### V. Generalised ontological model

Here we would like to generalise the above model by considering the space of processes. Denote $[\lambda]$ as a path in the ontic space. It could be a continuous path or a discrete, randomly jumping process. Two paths $[\lambda]_1$ and $[\lambda]_2$ are said to be *M*-equivalent if

$$p(k|[\lambda]_1, M) = p(k|[\lambda]_2, M), \quad \forall k, M$$

where $M$ is a quantum measurement and $k$ is an outcome of $M$.

Each set of these probabilities thus defines an equivalence class of paths. If a set of probabilities satisfy quantum probabilistic relations we call the paths Q-Equivalent.

Even though measurement outcomes may depend probabilistically on the paths in the above sense, a preparation processes (i.e. quantum states) in general need not correspond to distributions of processes, as is the case in the D-D model in section IV. In that model, there is no well-defined measure in the space of processes – the length of any process is arbitrary and is admissible as long as the relative frequencies in the path matches the quantum probabilities for any possible measurement.

But let's consider the special case where preparation do correspond to distributions over the space of processes, i.e. $p([\lambda]|\psi)$. This, together with measurement probabilities $p(k|[\lambda], M)$, constitutes a *Process-Based Probability-Probability Model* (a PPP model, in short). A PPP model is one in which preparation is a distribution over paths in the ontic state space, while measurement outcomes are probabilistically dependent on the paths. When the paths reduces to single points in ontic space, a PPP model reduces to the standard ontological models:

$$p([\lambda]|\psi) \to p(\lambda|\psi), \quad p(k|[\lambda], M) \to p(k|\lambda, M)$$

Therefore, in analogy of the PBR theorem, we can obtain a more general ontological theorem, one that is about processes underlying quantum states:

> **Ontological theorem about processes**
>
> *For an ontological model where the preparation and measurement processes are represented by distributions on the space of processes, $p([\lambda]|\psi)$ and $p(k|[\lambda], M)$, respectively, to reproduce quantum probability structure, the distributions $p([\lambda]|\psi)$ are non-overlapping.*

This theorem says, in order to be consistent with quantum theory, quantum states are ontic about the processes - knowing the process in which a system is undergoing determines the quantum state in which the system was prepared. This theorem reduces to the standard PBR theorem when the paths reduces to ontic states at a single moment.

Even though quantum state is ontic in this sense, it is easy to incorporate indistinguishability of non-orthogonal states in this generalized ontological model, as measurement here does not simply read out the ontic state of the system. For example, distinct non-orthogonal quantum states can correspond to processes that intersect with each other in the ontic space. They are indistinguishable if all measurement contain outcomes that are triggered by these intersecting points.

More precisely, the model can be based on the following definition of $p([\lambda]|\psi)$ and $p(k|[\lambda], M)$:

For non-orthogonal quantum states $\psi_1, \psi_2$, $[\lambda]_1$ and $[\lambda]_2$ are the paths followed by the system, respectively:

$$p([\lambda]_1|\psi_1) > 0, p([\lambda]_2|\psi_2) > 0$$

We assume that if a path contains some set of ontic states $\{\lambda\}_{E_k}$ it would induce an outcome $E_k$ for measurement $M$: $p(k|[\lambda], M) > 0$ if $\{\lambda\}_{E_k} \in [\lambda]$.

If for every measurement $M$, there is at least one outcome $E_k$ such that

$$[\lambda]_i \cap \{\lambda\}_{E_k} \neq \emptyset \qquad (i = 1, 2)$$

then $\psi_1$ and $\psi_2$ are indistinguishable. Note that $[\lambda]_1$ and $[\lambda]_2$ need not be intersecting at the points in $\{\lambda\}_{E_k}$.

Interestingly, intersecting paths for distinct non-orthogonal quantum states means that these quantum states correspond to the same ontic state at the intersection, thus being "epistemic" in the standard sense at some moment in time.

## VI. Conclusion

It is usually thought that for a realist to embrace the epistemic view of quantum state, the quantum state is to be understood as our knowledge about the underlying reality, in the form of distributions over the set of ontic states, $p(\lambda|\psi)$. On the other hand, proponents of $\psi$-epistemic view that do not take quantum states to represent our knowledge of the underlying reality are commonly viewed as non-realists. This view is dispelled by the simple model in section IV (see also first subsection of section V). The model is realist (as it proposes a set of ontic states) and yet the quantum states are epistemic not in the sense of representing knowledge about underlying ontic states, but representing our knowledge about measurement outcomes. Moreover, this knowledge is objective in the sense that the process determines the quantum state - anyone knowing about the process will assign the same probabilities to measurement outcomes, via the relative frequency rules.

We introduced two hybrid models (the P-D and D-P models) where either the preparation or the measurement is not a distribution over the ontic states. We also introduced a generalized class of ontological models (PPP model) where both preparation and measurement depend on distributions over processes, instead of distributions over ontic states. This model obeys the generalized ontological theorem about processes – i.e. the quantum states are "ontic" with respect to the processes, not with respect to the ontic states. This generalized ontological theorem reduces to the usual PBR theorem when the paths reduces to points in the ontic state space.

# References

[1] J. S. Bell, Physics, 1, 195 (1964).

[2] S. Kochen and E. Specker, Journal of Mathematics and Mechanics, 17, 59 (1967).

[3] N. Harrigan and R. W. Spekkens, Found. Phys. 40, 125 (2010), arXiv:0706.2661.

[4] M. F. Pusey, J. Barrett, and T. Rudolph, Nature Physics 8, 476 (2012), arXiv:1111.3328.

[5] M. S. Leifer, Quanta **3**, 68 (2014), arXiv:1409.1570

[6] R. W. Spekkens, Phys. Rev. A 75, 032110 (2007), arXiv:quant-ph/0401052.